\documentclass[11pt,twoside]{article}

\usepackage{ihep,epsfig,amsfonts,amsmath,amssymb,amsbsy}

\begin{document}%%%%%%%%%%%%%%%%%%%

\title{\textbf{\uppercase{Black hole mass decreasing due to~phantom energy accretion}}%
}
\author{\textbf{E.\,Babichev,%
\footnote{babichev@inr.npd.ac.ru}%
V.\,Dokuchaev%
\footnote{dokuchaev@inr.npd.ac.ru}
and %
Y.\,Eroshenko%
\footnote{erosh@inr.npd.ac.ru}
} \\[2mm]
\textit{Institute for High Energy Physics, 142281 Protvino,
Russia}}
\date{}
\maketitle

\begin{abstract}
Solution for a stationary spherically symmetric accretion of the
relativistic perfect fluid with an equation of state $p(\rho)$
onto the Schwarzschild black hole is presented. This solution is a
generalization of Michel solution and applicable to the problem of
dark energy accretion. It is shown that accretion of phantom
energy is accompanied with the gradual decrease of the black hole
mass. Masses of all black holes tend to zero in the phantom energy
universe approaching to the Big Rip.
\end{abstract}

\section{Introduction}

Our Universe is seems to undergo a period of accelerated expansion
and it is assumed that a considerable part of the total density
consists of dark energy component with negative 
pressure~\cite{acceler}. There are several candidates for the dark energy:
cosmological constant ($\Lambda$) or dynamical component such as
quintessence \cite{CaDaSt} and $k$-essence \cite{ArMuSt}. In
connection with the solving of the problem of fine-tuning the
models of dynamical dark energy component are seem to be more
realistic as they admit to construct ``tracker'' \cite{tracker} or
``attractor'' \cite{ArMuSt} solutions.

One of the peculiar feature of the cosmological dark energy is a
possibility of the Big Rip \cite{Caldw}:  the infinite expansion
of the universe during a finite time.  The Big Rip scenario is
realized if a dark energy is in the form of the phantom energy
with $\rho+p<0$. In this case the scenario of Big Rip is possible
when cosmological phantom energy density grows at large times and
disrupts finally all bounded objects up to subnuclear scale. Note,
however that the only condition $\rho+p<0$ is not enough for the
realization of Big Rip \cite{McInnes}. In \cite{Alam} the authors
analyzed the supernova data in the model independent manner and
showed that the presence of the phantom energy with $-1.2<w<-1$ is
preferable in the present moment of time. The analogy between
phantom and QFT in curved space-time has been developed in
\cite{NojOdi}. The entropy of the universe with phantom energy is
discussed in \cite{BreNojOdiVan}.

Usually the evolution of quintessence or $k$-essence are
considered in a view of cosmological problems. However in the
presence of compact objects such as black holes the evolution of
dark energy should be sufficiently different from that in the
cosmological consideration. Indeed, what would be the fate of
black holes in the universe filled with the phantom energy and
coming to Big Rip? Recently we showed that all black holes
gradually decrease their masses and very near the Big Rip they
finally disappear \cite{BDE}. In the present work we study in
details the stationary accretion of dynamical dark energy into the
black hole. As a model of DE we take the perfect fluid with
negative pressure. The studying of accretion of perfect fluid on
the compact objects originated from Bondi~\cite{Bon}. The
relativistic generalization of the perfect fluid accretion were
made by Michel~\cite{Mic}.  Below we find the solution for a
stationary accretion of the relativistic perfect fluid with an
arbitrary equation of state $p(\rho)$ onto the Schwarzschild black
hole. Using this solution we show that the black hole mass
diminishes by accretion of the phantom energy. Masses of all black
holes gradually tend to zero in the phantom energy universe
approaching to the Big Rip. The diminishing of a black hole mass
is caused by the violation of the energy domination condition
$\rho+p\geq0$ which is a principal assumption of the classical
black hole `non-diminishing' theorems \cite{hawkell}. The another
consequence of the existence of a phantom energy is a possibility
of traversable wormholes \cite{wormh}.
 In~\cite{Jac,BeaMag,FroKof,Unruh} authors studied the accretion of
scalar quintessence field into the black hole, using the specific
quintessence potentials $V(\phi)$ for the obtaining of the
analytical solution for the black hole mass evolution.  We use
essentially different approach for the description of DE accretion
into black hole, namely, we model the DE by the perfect fluid with
the negative pressure.

\section{General equations}
\label{Gen}

Let us consider the spherical accretion of dark energy onto black
hole. We assume that the density of the dark energy is
sufficiently low so that the metric can be described by
Schwarzschild metric. We model the dark energy by a perfect fluid
with  energy-momentum tensor: $T_{\mu\nu}=(\rho+p)u_\mu u_\nu -
pg_{\mu\nu}$, where $\rho$ is the density and $p$ is the pressure
of the dark energy and $u_{\mu}$ is the four-velocity
$u^\mu=dx^\mu/ds$. The integration of the time component of the
energy-momentum conservation law $T^{\mu\nu}_{\;\;\; ;\nu}=0$
gives the first integral of motion
\begin{equation}
 \label{eq1}
  (\rho+p)\left(1-\frac{2}{x}+u^2\right)^{1/2}x^2 u =C_1,
\end{equation}
where $x=r/M$, $u=dr/ds$ and $C_1$ is a constant determined below.

Given the equation of state $p=p(\rho)$, one can introduce the
function $n$ by the relation:
\begin{equation}
 \label{n}
 \frac{d\rho}{\rho+p}=\frac{d n}{n}.
\end{equation}
The function $n$ plays the role of concentration of the particles, though one
can use $n$ for the media without introducing any particles.  In this case $n$
is the auxiliary function.  For general equation of state $p=p(\rho)$, from
(\ref{n}) we obtain the following solution for $n$:
\begin{equation}
  \label{n1}
  \frac{n(\rho)}{n_\infty}=\exp\left(\,\,
  \int\limits_{\rho_{\infty}}^{\rho}\frac{d\rho'}{\rho'+p(\rho')}\right)\
  .
\end{equation}
From the conservation of energy-momentum along the velocity
$u_{\mu}T^{\mu\nu}_{\quad ;\nu}=0.$ using (\ref{n1}) we obtain the
another first integral:
\begin{equation}
  \label{flux}
  \frac{n(\rho)}{n_\infty}ux^2=-A,
\end{equation}
where $n_\infty$ (the concentration of the dark energy at the
infinity) was introduced for convenience. In the case of inflow
$u=(dr/ds)<0$ and the constant $A>0$. From (\ref{eq1}) and
(\ref{flux}) one can easily obtain:
\begin{equation}
  \label{energy}
  \frac{\rho+p}{n}\left(1-\frac{2}{x}+u^2\right)^{1/2} =C_2,
\end{equation}
where
\begin{equation}
\label{C2} C_2=\frac{\rho_\infty+p(\rho_\infty)}{n(\rho_\infty)} \ .
\end{equation}
We will see below that the constant $A$ which determines the flux
is fixed for fluids with \mbox{$\partial p/\partial\rho>0$.} This can be
done through finding of the critical point.  Following Michel
\cite{Mic} we obtain the para\-me\-ters of critical point:
\begin{equation}
\label{cpoint} u_*^2=\frac{1}{2 x_*},\quad
V_*^2=\frac{u_*^2}{1-3u_*^2},
\end{equation}
where
\begin{equation}
\label{V1} V^2=\frac{n}{\rho+p}\frac{d(\rho+p)}{dn} -1.
\end{equation}
From this by using (\ref{n}) it follows that $V^2=c_s^2(\rho)$,
where $c_s^2=\partial p/\partial\rho$ is the squared effective
speed of sound in the media.  Combining the Eqs. (\ref{energy}),
(\ref{C2}), (\ref{cpoint}) and (\ref{V1}) we find the following
relation:
\begin{equation}
 \label{rho_c}
 \frac{\rho_*+p(\rho_*)}{n(\rho_*)}=
 \left[1+3c_s^2(\rho_*)\right]^{1/2}
 \frac{\rho_\infty+p(\rho_\infty)}{n(\rho_\infty)},
\end{equation}
which gives the $\rho_*$ for arbitrary equation of state
$p=p(\rho)$.  Given $\rho_*$ one can find $n(\rho_*)$ 
using~(\ref{n1}) and values $x_*$, $u_*$, using (\ref{cpoint}) and
(\ref{V1}).  Then substituting the calculated values in
(\ref{flux}) one can find the constant $A$.  Note that there is no
critical point outside the black hole horizon ($x_*>1$) for
$c_s^2<0$ or $c_s^2>1$.  This means that for unstable perfect
fluid with $c_s^2<0$ or $c_s^2>1$ a dark energy flux onto the
black hole depends on the initial conditions. This result has a
simple physical interpretation: the accreting fluid has the
critical point if its velocity increases from subsonic to
trans-sonic values.  In a fluid with a negative $c_s^2$ or with
$c_s^2>1$ the fluid velocity never crosses such a point. It should
be stressed, however, that fluids with $c_s^2<0$ are
hydrodynamically unstable (see discussion in
\cite{FabMar97,Carroll}). The Eq. (\ref{energy}) together with
(\ref{n1}) and (\ref{flux}) describe the requested accretion flow
onto the black hole. These equations are valid for perfect fluid
with an arbitrary equation of state $p=p(\rho)$, in particular,
for a gas with zero-rest-mass particles (thermal radiation) and
for a gas with nonzero-rest-mass particles. For a
nonzero-rest-mass gas the couple of equations (\ref{flux}) and
(\ref{energy}) is reduced to similar ones found by Michel
\cite{Mic}. One would note that the set of equations (\ref{n1}),
(\ref{flux}) and (\ref{energy}) are also correct in the case of
dark energy and phantom energy $\rho+p<0$. In this case
concentration $n(\rho)$ is positive for any $\rho$ and constant
$C_2$ in (\ref{energy}) is negative.

The black hole mass changes at a rate $\dot M=-4\pi r^2T_0^{\;r}$
due to the fluid accretion. With the help of (\ref{flux}) and
(\ref{energy}) this can be expressed as
\begin{equation}
 \label{evol}
 \dot{M}=4\pi A M^2 [\rho_{\infty}+p(\rho_{\infty})].
\end{equation}
For the phantom energy the relation (\ref{evol}) leads to the
diminishing of the black hole mass. That means that in the
universe filled with phantom energy the black holes should melt
away. This result is general, it does not depend on the equation
of state $p=p(\rho)$, the only condition $p+\rho<0$ is important.

\section{The analytical models}

\label{Mod}

Let us consider the model of dark energy with linear dependence of
pressure from the density:
\begin{equation}
 \label{p1}
 p=\alpha(\rho - \rho_0),
\end{equation}
which include, among others, the ultra-relativistic gas
($p=\rho/3$) and simplest models of dark energy ($\rho_0=0$ and
$\alpha<0$). Introduced value $\alpha$ is connected with usual
equation of state $w=p/\rho$ by the relation
$w=\alpha(\rho-\rho_0)/\rho$. For $\alpha<0$ there is no critical
point for the flux of the fluid into the black hole. In the case
of $\alpha>0$, using (\ref{cpoint}) and (\ref{V1}) we find the
parameters for critical point in model (\ref{p1}):
\begin{equation}
\label{cpoint1} x_*=\frac{1+3\alpha}{2\alpha},\quad
u_*^2=\frac{\alpha}{1+3\alpha} \ .
\end{equation}
It should be noted that in the linear model (\ref{p1}) the
parameters of critical point (\ref{cpoint1}) determined only by
$\partial p/\partial\rho=\alpha$ and do not depend on the
parameter $\rho_0$, which determines what physical fluid is
considered:  relativistic gas, dark energy or phantom energy. Note
also that for $\alpha>1$ (that corresponds to the non-physical
situation of superluminal speed of sound) there is no critical
point outside the black hole. Let us calculate the constant $A$
which determines the flux of the fluid into the black hole. From
Eq. (\ref{n1}) we find:
\begin{equation}
  \label{n2}
  \frac{n}{n_\infty}=\left|\frac{\rho_{\rm
  eff}}{\rho_{\rm eff,\infty}}\right|^{1/(1+\alpha)},
\end{equation}
where we defined the effective density $\rho_{\rm eff}\equiv\rho+p
= -\rho_0\alpha +(1+\alpha)\rho$. Using (\ref{rho_c}) we obtain:
\begin{equation}
\label{rho_eff_c} \left(\frac{\rho_{\rm eff_*}}{\rho_{\rm
eff,\infty}}\right)^{\alpha/(1+\alpha)}= (1+3\alpha)^{1/2},
\end{equation}
where $\rho_{\rm eff_*}$ is the value of effective density at the
critical point and $\rho_{\rm eff,\infty}$ is the effective
density at the infinity.  Substituting (\ref{rho_eff_c}) in
(\ref{n2}) and then using (\ref{flux}) we find for linear model:
\begin{equation}
\label{A1}
A=\frac{(1+3\alpha)^{(1+3\alpha)/2\alpha}}{4\alpha^{3/2}} \ .
\end{equation}
It is easily seen that $A\ge 4$ for $0<\alpha<1$. For $\alpha=1$
that corresponds to $c_s=1$ the we have $A=4$. From this we may
conclude that for typical sound speeds the constant $A$ has value
around unity. For some particular choices of parameter $\alpha$
the values $\rho(x)$ and $u(x)$ can be calculated analytically.
For example, for $\alpha=1/3$ the fluid density is given by:
\begin{equation}
 \label{sol1}
 \rho=\frac{\rho_0}{4}+\left(\rho_{\infty}-\frac{\rho_0}{4}\right)
 \left[z+\frac{1}{3(1-2x^{-1})}\right]^2,
\end{equation}
where
$$
 z=\left\{ \begin{array}{ll}
 -2{\sqrt{\frac{a}{3}}}\,\cos\left(\frac{2\,\pi }{3}
 -\frac{\beta}{3}\right),& 2<x<3,\\
 2{\sqrt{\frac{a}{3}}}\,\cos\left(\frac{\beta}{3}\right),& x>3,
\end{array} \right.
$$
$$
 \beta=\cos\left[\frac{b}{2\,(a/3)^{3/2}}\right]
$$
and
$$
 a=\frac{1}{3{\left( 1 - \frac{2}{x}\right) }^2}, \quad
 b=\frac{2}{27{\left(1-\frac{2}{x}\right) }^3}-
  \frac{108}{\left(1-\frac{2}{x}\right) x^4}.
$$
The density distribution for another physically interesting case
$\alpha=1$ is given by:
\begin{equation}
 \label{sol2}
 \rho=\frac{\rho_0}{4}+\left(\rho_{\infty}-\frac{\rho_0}{4}\right)
 \left(1+\frac{2}{x}\right)\left(1+\frac{4}{x^2}\right).
\end{equation}
The corresponding radial fluid velocity $u=u(x)$ can be calculated
by substituting of (\ref{sol1}) or (\ref{sol2}) into (\ref{eq1}).
For $\rho_0=0$ the solutions (\ref{sol1}) and (\ref{sol2})
describe correspondingly a thermal radiation and a fluid with
ultra-hard equation of state. In the case of
$\rho_\infty<\alpha\rho_0/(1+\alpha)$ the solutions (\ref{sol1})
and (\ref{sol2}) describe the phantom energy  falling onto the
black hole. For example, a phantom energy flow with parameters
$\alpha=1$ and $\rho_0= 4\rho_\infty$ results in a black hole mass
diminishing with the rate $\dot M=-8\pi(2 M)^2\rho_\infty$.

\section{Black holes in the  universe with Big Rip}
\label{Cosm}

Now we turn to the problem of the black hole evolution in the
universe with the Big Rip when a scale factor $a(t)$ diverges at
finite time \cite{Caldw}. For simplicity we will take into account
only dark energy and will disregard all others forms of energy.
The Big Rip solution is realized for in the linear 
model~(\ref{p1}) for $\rho+p<0$ and $\alpha<-1$. From the Friedman
equations for the linear equation of state model one can obtain:
$|\rho+p|\propto a^{-3(1+\alpha)}$. Taking for simplicity
$\rho_0=0$ we find the evolution of the density of a phantom
energy in the universe:
\begin{equation}
\label{sol3}
\rho_\infty=\rho_{\infty,i}\left(1-\frac{t}{\tau}\right)^{-2},
\end{equation}
where
\begin{equation}
\label{tau2}
\tau^{-1}=-\frac{3(1+\alpha)}{2}\left(\frac{8\pi}{3}
\rho_{\infty,i}\right)^{1/2}
\end{equation}
and $\rho_{\infty,i}$ is the initial density of the cosmological
phantom energy and the initial moment of time is chosen so that
the `doomsday' comes at time $\tau$. From (\ref{sol3}) and
(\ref{tau2}) it is easy to see that the Big Rip solutionis
realized for $\alpha\equiv\partial p/\partial\rho<-1$. In general,
the satisfying the condition $\rho+p<0$ is not enough for the
possibility for Universe to come to Big Rip. From (\ref{evol})
using (\ref{sol3}) we find the black hole mass evolution in the
universe coming to the Big Rip:
\begin{equation}
 \label{mevol1}
 M=M_i\left(1+\frac{M_i}{\dot M_0 \;\tau}\;
 \frac{t}{\tau-t}\right)^{-1},
\end{equation}
where
\begin{equation}
\dot M_0=(3/2)\,A^{-1}|1+\alpha|,
\end{equation}
and $M_i$ is the initial mass of the black hole. For $\alpha=-2$
and typical value of $A=4$ (corresponding to $u_{\rm H}=-1$) we
have $\dot M_0=3/8$. In the limit $t\to\tau$ (i.e. near the Big
Rip) the dependence of black hole mass on $t$ becomes linear,
$M\simeq\dot M_0\,(\tau-t)$. While $t$ approaches to $\tau$ the
rate of black hole mass decrease does not depend on both an
initial black hole mass and the density of the phantom energy:
$\dot M\simeq-\dot M_0$. In other words masses of all black holes
in the universe tend to be equal near the Big Rip. This means that
the phantom energy accretion prevails over the Hawking radiation
until the mass of black hole is the Planck mass. However, formally
all black holes in the universe evaporate completely at Planck
time before the Big Rip due to Hawking radiation.

\section{Scalar field accretion}

\label{Scalar}

In remaining let us confront our results with the calculations of
(not phantom) scalar field accretion onto the black hole
\cite{Jac,BeaMag,FroKof,Unruh}. The dark energy is usually
modelled by a scalar field $\phi$ with potential $V(\phi)$. The
perfect fluid approach is more rough because for given `perfect
fluid variables' $\rho$ and $p$ one can not restore the `scalar
field variables' $\phi$ and $\nabla\phi$. In spite of the pointed
difference between a scalar field and a perfect fluid we show
below that our results are in a very good agreement with the
corresponding calculations of a scalar field accretion onto the
black hole.

The Lagrangian of a scalar field is $L=K-V$, where $K$ is a
kinetic term of a scalar field $\phi$ and $V$ is a potential. For
the standard choice of a kinetic term $K=\phi_{;\mu}\phi^{;\mu}/2$
the energy flux is $T_{0r}=\phi_{,t}\phi_{,r}$. Jacobson
\cite{Jac} found the scalar field solution in Schwarzschild metric
for the case of zero potential $V=0$:
$\phi=\dot\phi_\infty[t+2M\ln(1-2M/r)]$, where $\phi_\infty$ is
the value of the scalar field at the infinity. In \cite{FroKof} it
was shown that this solution remains valid also for a rather
general form of runaway potential $V(\phi)$.  For this solution we
have $T_0^{\;r}=-(2M)^2\dot\phi^2_\infty/r^2$ and correspondingly
$\dot M=4\pi(2M)^2{\dot{\phi}}^2_\infty$.

The energy-momentum tensor constructed from Jacobson solution
completely coincides with one for perfect fluid in the case of
ultra-hard equation of state $p=\rho$ under the replacement
$p_{\infty}\to\dot{\phi}_\infty^2/2$,
$\rho_{\infty}\to\dot{\phi}_\infty^2/2$. It is not surprising
because the theory of a scalar field with zero potential $V(\phi)$
is identical to perfect fluid consideration \cite{Luk80}. In a
view of this coincidence it is easily to see the agreement of our
result (\ref{evol}) for $\dot M$ in the case of $p=\rho$ and the
corresponding result of \cite{Jac,FroKof}.

To describe the phantom energy the Lagrangian of a scalar field
must have a negative kinetic term~\cite{Caldw}, for example,
$K=-\phi_{;\mu}\phi^{;\mu}/2$ (for the more general case of the
negative kinetic term see~\cite{Gonz}). In this case the phantom
energy flux onto black hole has the opposite sign,
$T_{0r}=-\phi_{,t}\phi_{,r}$, where $\phi$ is the solution of the
same Klein-Gordon equation as in the case of standard scalar
field, however with the replacement $V\to-V$. For zero potential
this solution coincides with that obtained by Jacobson \cite{Jac}
for a scalar field with the positive kinetic term. Lagrangian with
negative kinetic term and $V(\phi)=0$ does not describe, however,
the phantom energy. At the same time, the solution for scalar
field with potential $V(\phi)=0$ is the same as with a positive
constant potential $V_0=const$, which can be chosen so that
$\rho=-\dot\phi^2/2+V_0>0$. In this case the scalar field
represents the required accreting phantom energy $\rho>0$ and
$p<-\rho$ and provides the decrease of black hole mass with the
rate $\dot M= -4\pi(2M)^2{\dot{\phi}}^2_\infty$.

The simple example of phantom cosmology (without a Big Rip) is
realized for a scalar field with the potential $V=m^2\phi^2/2$,
where $m\sim10^{-33}$~eV \cite{SamTop}. After short transition
phase this cosmological model tends to the asymptotic state with
$H\simeq m\phi/3^{1/2}$ and $\dot\phi\simeq2m/3^{1/2}$.  In the
Klein-Gordon equation the $m^2$ term (with the mentioned
replacement $V\to-V$) is comparable to other terms only at the
cosmological horizon distance. This means that the Jacobson
solution is valid for this case also. Calculating the
corresponding energy flux one can easily obtain $\dot
M=-4\pi(2M)^2\dot\phi^2_{\infty} =-64M^2m^2/3$. For
$M_0=M_{\odot}$ and $m=10^{-33}$~eV the effective time of black
hole mass decrease is $\tau=(3/64)M^{-1}m^{-2}\sim 10^{32}$~yr.\\[-2mm]

This work was supported in part by the Russian Foundation for
Basic Research grants 02-02-16762-a, 03-02-16436-a and
04-02-16757-a and the Russian Ministry of Science grants
1782.2003.2 and 2063.2003.2.

\end{document}